\documentclass[prl,amsmath,amsfonts,amssymb,superscriptaddress,showpacs]{revtex4}

\usepackage{graphicx,psfrag}
\usepackage{dcolumn}
\usepackage{bm}
\usepackage{mathbbol}
\usepackage{makeidx}               
\usepackage{graphicx}              
\usepackage[bookmarksnumbered,colorlinks,plainpages,citecolor=blue,urlcolor=blue]{hyperref}
\usepackage[rflt]{floatflt}
\usepackage{subfigure}
\usepackage{caption2}
\usepackage{latexsym}
\usepackage{amsmath}
\usepackage{amsfonts}
\usepackage{amssymb}
\usepackage{enumerate}

\newcommand{\uck}[1]{\o}

\def\beq{\begin{equation}}
\def\eeq{\end{equation}}
\def\bea{\begin{eqnarray}}
\def\eea{\end{eqnarray}}

\begin{document}

\begin{titlepage}
\date{\today}
\title{A Current of the Cheshire Cat's Smile: Dynamical Analysis of Weak Values}

\author{Yakir Aharonov}
\email{yakir@post.tau.ac.il}
\affiliation{\mbox{School of Physics and Astronomy, Tel Aviv University, Tel Aviv  6997801, Israel}
\mbox{Schmid College of Science, Chapman University, Orange, CA 92866, USA,}}

\author{Eliahu Cohen}
\email{eliahuco@post.tau.ac.il}
\affiliation{\mbox{H.H. Wills Physics Laboratory, University of Bristol, Tyndall Avenue, Bristol, BS8 1TL, U.K}}

\author{Sandu Popescu}
\email{s.Popescu@bristol.ac.uk}
\affiliation{\mbox{H.H. Wills Physics Laboratory, University of Bristol, Tyndall Avenue, Bristol, BS8 1TL, U.K}}


\maketitle

\section{Abstract}

Recently it was demonstrated, both theoretically and experimentally, how to separate a particle from its spin, or any other property, a phenomenon known as the ``Quantum Cheshire Cat''.  We present two novel gedanken experiments, based on the quantum Zeno effect, suggesting a dynamical process thorough which this curious phenomenon occurs. We analyze, for the first time, a quantum current consisting of spin without mass. Thus, the quantum variables of pre- and post-selected particles are understood to be involved in various interactions, even in the absence of their owners. This current is shown to provide a local explanation for seemingly nonlocal interactions.

\section{I. Introduction}

The Cheshire Cat experiment \cite{Cheshire1,Cheshire2} demonstrates a unique property of pre- and post-selected systems: we can find the spin of a particle (or polarization of a photon) even in places where the particle seems to be absent. The same can be generalized to every property of the quantum particles.

This observation was possible, {\it in a non-counterfactual way}, by performing weak measurements \cite{AAV,ACE} of the particle's spin and position, and finding their so called ``weak values''. Weak values can be seen as a generalization of expectation values to cases where the measured system has two different boundary conditions - at some initial and some final times. Within the Two-State Vector Formalism (TSVF) of quantum mechanics, these boundary conditions, known as pre- and post-selection, can be understood to evolve forward and backward in time (respectively) to define a richer notion of quantum reality at the intermediate times \cite{AV,Collapse}. The weak values are defined at any moment $t$ by

\begin{equation} \label{WV}
\langle A\rangle_w (t) = \frac{\langle \phi(t) | A | \psi(t) \rangle}{\langle \phi | \psi \rangle},
\end{equation}
where $\phi$ and $\psi$ are the pre-/post-selected states respectively, and $A$ is the measured operator.

The formulation of the original Cheshire Cat thought experiment is as follows:
Let a particle have two degrees of freedom: spatial $|L\rangle$,$|R\rangle$ and spinorial $|\uparrow\rangle$,$|\downarrow\rangle$ along the $z$ direction. The particle is pre-selected at the initial time in the entangled state
\begin{equation}
|\psi_i\rangle=\frac{1}{2}(|\uparrow\rangle+|\downarrow\rangle)|L\rangle+ \frac{1}{2}(|\uparrow\rangle-|\downarrow\rangle) |R\rangle.
\end{equation}

Later on, the particle is post-selected (using a projective measurement) in the product state
\begin{equation}
|\psi_f\rangle=\frac{1}{2}(|\uparrow\rangle-|\downarrow\rangle) (|L\rangle+|R\rangle).
\end{equation}

Using weak measurements at intermediate times, the particle is understood to be on the {\it right}, since
\begin{equation}
\langle \Pi_L \rangle_w=0~,~\langle \Pi_R \rangle_w=1,
\end{equation}
and it has an ``up'' spin along the $z$ direction, because
\begin{equation}
\langle \sigma_z \rangle_w =1.
\end{equation}
However, its spin is on the {\it left}
\begin{equation}
\langle \sigma_z \Pi_L \rangle_w=1~,~\langle \sigma_z \Pi_R \rangle_w=0.
\end{equation}
This can be understood as the failure of the product rule for weak values between pre- and post-selected states: the weak value of the product does not equal the product of weak values.

To better monitor this process we shall now slightly modify it to include the quantum Zeno effect \cite{Zeno}.

In the rest of  of the paper we will propose a dynamic analysis of this phenomenon, in a manner which will allow us to deduce some surprising features of pre- and post-selected ensembles. More generally, we will be able to understand quantum nonlocality, in some cases, as a local phenomenon of current without mass.

\section{II. Dynamical Separation of a Particle and its Spin}
In what follows we shall demonstrate the gradual (local) process resulting in the fading cat, and the ``massless current''. Throughout the calculation $\hbar=1.$ \hfill\break
Consider an electron free to move inside a $1D$ cavity. The left wall, located at $x=-L$, is a perfect reflector and the right wall at $x=L$ has the following interaction with the electron's spin:
\begin{equation} \label{H_int}
H_{int}=\frac{1+\sigma_z}{2}V_0\Theta(x-L),
\end{equation}
where $\Theta(x)=1/2+x/2|x|$ is the Heaviside function, i.e. the electron is facing a high step potential only if it has spin up along the $z$-direction.
Between the walls, at $x=0$, a high, narrow potential barrier is placed, with a very small transmission coefficient $\sin^2\alpha$, where $\alpha<<1$ (see Fig. \ref{Fig1}). This potential exerts the quantum Zeno effect which is essential to our analysis.
The electron is prepared in a well localized state on the left side of the cavity with spin up along the $x$ direction
\begin{equation} \label{initial}
|\psi\rangle=|L\rangle|\sigma_x=+1\rangle.
\end{equation}
Let the electron also have a momentum $\Delta p<<p<<\sqrt{2mV_0}$ to the right. When hitting the narrow potential its state will change according to:
\begin{equation}
\begin{array} {lcl}
|L\rangle \rightarrow \cos\alpha|L\rangle+\sin\alpha|R\rangle, \\
|R\rangle \rightarrow -\sin\alpha|L\rangle+\cos\alpha|R\rangle.
\end{array}
\end{equation}

The $|\sigma_z=+1\rangle$ component of the electron's spin interacts strongly with the left wall.
At times $\frac{2mL}{p}, \frac{4mL}{p},..., \frac{2mnL}{p}\equiv n\tau $, where $n$ is some integer, the electron has an approximate probability of $\sin^2\alpha$ to leave the left side of the cavity and tunnel to the right. These small amplitudes are coherently added at the right side of the cavity, hence after $n$ such cycles, the electron's state will be:
\begin{equation}
|\psi (n\tau) \rangle=\cos(n\alpha)|L\rangle|\sigma_z=+1\rangle+\sin(n\alpha)|R\rangle|\sigma_z=+1\rangle.
\end{equation}
Therefore, when $n\alpha\approx\frac{\pi}{2}$, i.e. $n \propto 1/\alpha$, we have a non-negligible probability to find the electron on the right side of the cavity.

However, if the electron has spin $|\sigma_z=-1\rangle$, fractions of $\sin^2\alpha$ ``leak'' after times $\frac{mL}{p}, \frac{3mL}{p},...,\frac{(2n+1)mL}{p}$. In this case, the amplitudes are not-coherently added, hence after $1/\alpha$ cycles, there is still a high probability to find the electron on the right side:
\begin{equation}
(\cos \alpha)^{\frac{2}{\alpha}}\approx (1-\frac{\alpha^2}{2})^{2/\alpha} \approx 1 - \alpha \approx 1.
\end{equation}

For this reason, if after time $T=\pi\tau/2\alpha$ we find the electron in the left side of the cavity, we conclude that most-likely it has spin $|\sigma_z=-1\rangle$ (note that as opposed to the original Cheshire Cat experiment, post-selection is now performed only on the spatial degree of freedom, which nevertheless reveals the spin). In other words, if after time $T$ we measure its spin along the $x$-axis, there is approximately $50 \% $ chance we would find $|\sigma_x=-1\rangle$, that is, orthogonal to its initial state (Eq. \ref{initial}). It seems paradoxical that a particle changes its initial state to an orthogonal state, due to a force it rarely ``feels''.

How could this happen? During the time interval $(0,T)$ there was only a negligible probability of finding the electron on the right side of the cavity, but after time $T$ we find the effect of the potential in the right side with probability approaching $0.5$.

This apparent paradox can be resolved using the concept of the ``Quantum Cheshire Cat''. Indeed, if the spin of the electron could leave its mass behind (at the left side of the cavity) and travel to the right, we could understand this weird interaction in a local sense. Hence, the apparently nonlocal interaction is made local by the weak current of the spin. By using weak measurements at the intermediate times we will now test this claim.

\begin{figure}[h]
 \begin{center} \includegraphics[height=5cm]{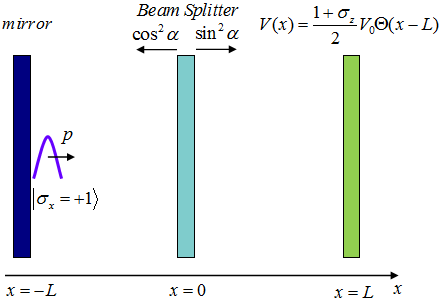}
      \caption{{\bf The Initial setup.} An electron is confined to a cavity whose right wall is spin-dependent. In the middle of the cavity there is a narrow potential with very low transmission coefficient.} \label{Fig1}
\end{center}
\end{figure}

The initial (pre-selected) wavefunction in Eq. \ref{initial} evolves to:
\begin{equation} \label{evo1}
|\psi (n\tau) \rangle=\frac{1}{\sqrt{2}}[\cos\alpha|L\rangle|\sigma_z=-1\rangle+\sin\alpha|R\rangle|\sigma_z=-1\rangle+
\cos(n\alpha)|L\rangle|\sigma_z=+1\rangle+\sin(n\alpha)|R\rangle|\sigma_z=+1\rangle].
\end{equation}
In addition, finding the electron on the left side of the cavity after time $T=\pi\tau/2\alpha$ is amount to a post-selection of the state:
\begin{equation}
\langle\phi|=\langle L|\langle\sigma_z=-1|,
\end{equation}
that is, although a position measurement was performed, the spin of the electron is also known. The backward evolution of this state is:
\begin{equation} \label{evo2}
\langle\phi (n\tau)|=\langle L|\langle\sigma_z=-1|\cos\alpha+\langle R|\langle\sigma_z=-1|\sin\alpha.
\end{equation}

For all intermediate times it can be easily found according to Eqs. \ref{WV},\ref{evo1},\ref{evo2}, that:
\begin{equation}
\langle \pi_R \sigma_z\rangle_w = -sin^2\alpha,~\langle \pi_L \sigma_z\rangle_w = -cos^2\alpha,
\end{equation}
where $\pi_L=|L\rangle \langle L|$ and $\pi_R=|R\rangle \langle R|$ are projections on the left and right sides of the cavity, respectively. Consequently, $\langle \sigma_z\rangle_w = -1$, which ought to be like that, because the post-selected state is an eigenstate of $\sigma_z$ with eigenvalue $-1$.

More importantly, the value of $\sigma_x$ on the right side of the cavity is given by the weak value (see Eq. \ref{WV})
\begin{equation}
\begin{array} {lcl}
\langle \pi_R \sigma_x(n\tau)\rangle_w = \sin \alpha \sin(n\alpha),~\langle \pi_L \sigma_x(n\tau)\rangle_w = \cos \alpha \cos(n\alpha).
\end{array}
\end{equation}
That is, $\sigma_x$ gradually increases over time on the right side of the cavity, despite the fact that the electron rarely reaches there. To find the total spin in the right side of the cavity we will sum over all the contributions until time $T=\frac{\pi \tau}{2\alpha}$:
\begin{equation}
\langle \pi_R \sigma_x^{tot}\rangle_w=\sum_{n=1}^{\lfloor \pi/2\alpha \rfloor} \sin \alpha \sin(n\alpha) \simeq
\int_0^{\pi/2\alpha} \sin \alpha \sin(n\alpha) dn = \frac{\sin\alpha}{\alpha} \simeq 1.
\end{equation}

To understand how $\sigma_x$ of the initial state changes over time we shall use the Heisenberg equation:
\begin{equation}
\frac{d}{dt}\sigma_x(t)=\frac{i}{\hbar}[H,\sigma_x]=-V_0\Theta(x)\sigma_y.
\end{equation}
The value of $\sigma_y$ between the pre- and post-selection in the right side of the cavity is given by the following weak value
\begin{equation}
\langle \pi_R \sigma_y(n\tau)\rangle_w = - i \sin \alpha \sin(n\alpha),
\end{equation}
where we used Eqs. \ref{evo1},\ref{evo2}. Hence, for $0<x<L$ there is an imaginary current of $\sigma_x$ accounting for the local change of the spin from $\sigma_x=+1$ at the beginning to $\sigma_x=-1$ at the end (to better understand the appearance of imaginary weak value here, see \cite{AC}). For $x\ge L$ there is a real, tunneling current of the $\sigma_z=+1$, as well as a real current of the transmitted $\sigma_z=-1$.


\section{III. Entangled Current}
We assume now that besides the electron, the right wall in Sec. II also has an inner degree of freedom (``spin'').  The interaction Hamiltonian would be now:
\begin{equation}
H_{int}=\frac{1+\sigma_z^{(1)}\sigma_z^{(2)}}{2}V_0\Theta(x-L),
\end{equation}
where $\sigma_z^{(1)}$ refers to the electron and $\sigma_z^{(2)}$ refers to the right wall, i.e. the electron interacts with wall only when there spins along the z-direction are parallel.


Assume the initial state of the system is:
\begin{equation}
|\psi\rangle=|L\rangle|\sigma_x^{(1)}=+1\rangle|\sigma_x^{(2)}=+1\rangle.
\end{equation}

Finding the electron after time $T$ in the left side of the cavity, singles out the states $|\sigma_z^{(1)}=+1\rangle |\sigma_z^{(2)}=+1\rangle$ and $|\sigma_z^{(1)}=-1\rangle |\sigma_z^{(2)}=-1\rangle$ of the two spins. Therefore, during the time evolution of the system, the spins of the electron and the wall have become maximally entangled, that is, their state have changed to
\begin{equation}
|\phi\rangle=|L\rangle\frac{\sigma_z^{(1)}=+1\rangle|\sigma_z^{(2)}=-1\rangle + |\sigma_z^{(1)}=-1\rangle|\sigma_z^{(2)}=+1\rangle}{\sqrt 2}
\end{equation}

In the previous section we saw a ``leakage'' of a one-particle-state, but now we observe a gradual leakage of an entangled state. As done before, we can monitor the process using weak measurements and verify the interaction at the meeting point after each period time. Thus, the seemingly nonlocal effect turns out to have a local description based on a massless current of entangled spins.

\section{IV. Discussion}
This work presents, for the first time, a dynamical analysis of weak values which sheds light on the Cheshire cat paradox and on the quantum Zeno effect. The spin of a particle, was shown to gradually flow from its mass while interacting with a potential barrier. It was then demonstrated how to entangle two distant systems by utilizing a current of entangled states. All the above processes, which at first seem nonlocal, can be understood locally using currents of spin weak values without mass. Similarly to the demonstration of interaction-free measurement with Zeno effect \cite{Kwiat}, the above predictions can be tested in a laboratory experiment with photons instead of electrons and a polarizing beam-splitter instead of the right wall.
\normalsize{\section{Acknowledgements}}
We acknowledge support of the Israel Science Foundation Grant No. 1311/14, of the ICORE Excellence Center ``Circle of Light'', of DIP, the German-Israeli Project cooperation and ERC AdG NLST.

\normalsize{

\end{titlepage}
\end{document}